\documentclass[prb,onecolumn,superscriptaddress]{revtex4}
\usepackage{amsmath,amssymb,graphicx,epsfig,color, float}
\usepackage{epstopdf}

\begin{document}
\title{Bibliography of Literature on GW Ab Initio Calculations which use Imaginary Time}

\author{Vincent Sacksteder IV}
\email{vincent@sacksteder.com}
\affiliation{Department of Physics, Wittenberg University, Ohio, USA}
\affiliation{Department of Physics, Royal Holloway University of London, UK}

\date{\today}

\begin{abstract}
The GW Approximation is an ab initio approach to calculating electronic structure which avoids using the Local Density (LDA) Approximation, the Generalized Gradient  (GGA) Approximation, or similar density functionals.  It goes beyond the Hartree-Fock approximation by including screening and excited state effects, and shares conceptual similarities with MP2 and RPA calculations.   Because GW includes dynamics and time/frequency dependence of the system's screening and excited state behavior, a pivotal issue in any GW calculation is the question of how to numerically represent  and manipulate time/frequency dependence.  While earlier GW calculations generally used a representation in real time/frequency, many recent calculations have used a representation on the imaginary time axis.  Imaginary time  is  important not only for numerics but also because it can enable additional physics such as systems where self-consistent GW is needed or that are strongly interacting, integration with DMFT, and scaling of GW to very large systems.  This current text reviews the research literature on imaginary time GW codes and briefly discusses the possibilities for memory and CPU optimizations.
\end{abstract}


\maketitle

This document is the result of  a thorough literature search which I performed in November 2016, finding all research articles which used imaginary time representations as a part of  GW approximation calculation of ab initio electronic structure.  The imaginary time representation is promising for five reasons:  
\begin{itemize}
\item Firstly, it offers a systematic solution for the divergences (poles) which occur in real time representations and for the numerical sensitivity that results from those poles. This solution is an important aid in doing self-consistent GW calculations which go beyond one-shot calculations.  \item Secondly, because imaginary time offers a systematic solution for poles, it is one of the most important ways of doing self-consistent GW calculations. The other major way to do self-consistent GW is Kotani and Schilfgaarde's quasi-particle qsGW.   My literature search was partly motivated by the desire to do self-consistent GW, and therefore it also reports two research efforts which report self-consistent GW results but used neither qsGW nor imaginary time.
\item Thirdly, the imaginary time representation can offer enormous increases in  CPU efficiency and scaling, as seen in the space-time approach pioneered by Rex Godby et al and recently implemented by the VASP code.  Using the space-time approach, GW calculations of many tens of atoms are possible on a small cluster - see recent VASP reports.  
\item Fourthly, since DMFT calculations are performed in imaginary time representations, and since there is ongoing work on combining GW (or constrained RPA) with DMFT to produce a fully ab initio strongly correlated code, it is desirable to use the same imaginary  time representation in the GW calculation that is used in the DMFT calculation.
\item Fifthly, imaginary time allows temperature to be included in an ab initio calculation in a complete self-consistent way.
\end{itemize}

I also include a brief enumeration and discussion of the possible memory and CPU optimizations that can be explored for GW codes.

 I hope that making these results public will stimulate further research on imaginary time methods,  encourage researchers who are considering whether to pursue this direction, and give a quick overview of the research already done and algorithms already used.  In the five years from 2011 to 2016 the number of imaginary time GW codes doubled.  The number of publicly available general purpose codes went from zero to two, with promises from VASP that they will soon distribute their code, making it the third one that is publicly available.  This is coupled to a much larger trend within the ab initio community toward MP2/RPA style calculations, and also  to a trend toward strongly interacting systems.

\section{List of self-consistent  or imaginary time GW calculations}
Bear in mind that I'm leaving out codes that do not use imaginary time representations, so if you don't find your favorite code here, that might be why.  If you find an omission, or you find that my description is incomplete or inaccurate, please don't get angry, just let me know, and I'll try to update this document when I find time.  More details of what I didn't include in this list are supplied below.
\begin{itemize}
\item Plane-wave codes:
\begin{itemize}
\item The New VASP MP2/RPA/GW code, 3 PRBs, 2014-2016. \cite{kaltak2014low, PhysRevB.90.054115,PhysRevB.94.165109} This code is much different from the previous VASP RPA code, which did not use imaginary times.  The new GW code does not yet include potentially risky features like extrapolations in basis size.  Many different insulators (C, BN, MgO, and LiF), semiconductors (Si, GaAs, SiC, and ZnO), metals (Cu and SrVO3).  Large supercells with up to 50 atoms. Plane wave basis augmented with PAW, and the articles explain how this is done in detail. 20 time/energy points. $8^3$ kpoints.  20 meV accuracy except for the metals.  Diamond with $4^3$ kpoints done in one hour on 64 cores. Many other examples running in one to four hours.  This code has also published calculations of forces, which requires very rigorous control of basis truncation and numerical  errors, and is unusual in a GW code.  Forces are important for structural optimization and structural optimization.
\item Lin-Wang Wang (Lawrence Berkeley National Laboratory), PRB, 2015. \cite{PhysRevB.91.125135} Si3, C3, O3, Al2, SiH2, HNO, CHF.  Does not take advantage of lattice momentum/position to reduce the size of matrices. Plane wave basis, norm-conserving pseudopotentials, 8000 Gvectors or 60000 points in real space.  400 time points for the Green's functions, 100 time points for the susceptibility and screened potential which decay more quickly.
\item Godby's space-time method, 1995-2000. \cite{PhysRevLett.74.1827,rieger1999gw,Steinbeck2000} Si, Fe, GaN. Si supercells with 8 atoms. Plane waves. 60-120 time/energy grid points, or only 15 grid points in the later article on an optimized space-time code. 100 meV accuracy.
\end{itemize}
\item Codes which seem to be oriented toward general purpose reliable usage by a possibly large audience:
\begin{itemize}
\item VASP, as above.
\item The ELK code, PRB, 2016.  \cite{PhysRevB.93.125210} 18 semiconductors and insulators, FP-LAPW basis. Uses a "uniform power mesh" from Ku and Eguiluz for time points, with 91 time points.
\item FHI-AIMS, many articles 2012-2015, many citations.  \cite{PhysRevB.80.045402,van2015gw, PhysRevB.92.081104,PhysRevLett.110.146403,PhysRevB.88.035120,PhysRevB.88.075105,PhysRevB.86.081102,paier2012assessment} 100+ molecules.  Atom centered orbitals.  Overlap matrices because the basis is non-orthogonal; explanation in the articles. Up to 60 time/energy grid points. 
\item Kutepov + Haule + Savrasov + Kotliar. (this may be private, but really impressive results). 1 PRL and 2 PRBs, 2006-2012. \cite{PhysRevB.85.155129,PhysRevB.80.041103,PhysRevLett.96.226403} FLAPW, muffin tins.  Na, Al, Si, Pu, Am, diamond, SiC, GaAs, ZnS, ZSe, Na, Ti, Ni, Fe, Gd.  30-60 energy points, with a rather complicated way of transforming between time and energy. Overlap matrices because the basis is non-orthogonal; explanation in the articles.
\end{itemize}
\item Other codes which seem to have more of a research orientation, with a limited number of users:
\begin{itemize}
\item Schilfgaarde + Choi-Kutepov-Haule-Kotliar. 2 arxiv articles 2015-2016. \cite{choi2015first,sponza2016non} Quasiparticle GW, not the self-consistent GW done by the others on this list.  2000 energy grid points.
\item Koval + Foerster + Sanchez Portal. (on top of SIESTA or NWCHEM.) 2011-2014. \cite{PhysRevB.89.155417,foerster2011n3} Gaussian basis sets. 16 small molecules. Spectral function approach to energy dependence.  50-500 energy points.  They never transform to from energy to time; not an imaginary time approach.  I included this because they do do self-consistent GW.
\item Rostgaard and Thygesen et al, 2 PRBs, 2010-2011. \cite{PhysRevB.81.085103,PhysRevB.83.115108}  Conduction through molecular junctions. PAW, Wannier functions, numerical atomic orbitals. 34 molecules. 40,000 time points.
\item Stan + Dahlen + van Leeuwen.  3 articles, 2005-2009. \cite{stan2006fully,dahlen2005self,stan2009levels} Slater function basis.  He, Be, Ne, Mg, H2, LiH.  Uses a "uniform power mesh" based on Ku and Eguiluz for time points.
\item Zein and Antropov, PRL, 2002. \cite{PhysRevLett.89.126402} Real not imaginary time, Fe, Ni. LMTO basis.  I included this because they do do self-consistent GW.
\item Ku and Eguiluz. 2 PRLs, another article, and a thesis, 1997-2002. \cite{PhysRevLett.89.126401,PhysRevLett.81.1662,eguiluz1998electronic} FLAPW. Uses a "uniform power mesh" for time points.
\end{itemize}
\item Not included in this list:
\begin{itemize}
\item We are not listing quasiparticle GW methods unless they use imaginary time/energy.
\item We are not listing works on the homogenous electron gas.
\item We are not listing works on GW with tight binding  models, models derived using Wannier functions, 1-D model systems, and other model systems, including some works on GW+DMFT.
\item We are not not mentioning the huge variety of chemistry codes which are trying to access the same physics (MP2 and RPA research and codes). In particular, there is a huge literature on use of the Laplace transform to evaluate MP2 energies, which is essentially use of an imaginary time formulation.
\end{itemize}
\end{itemize}

\section{Basic Optimizations}
The most basic optimization steps available to a GW code are:
\begin{enumerate} 
\item  \textbf{The Time/Energy Grids.} Finding an efficient representation for the nonuniform time/energy grids, and for transforming between them, is a problem held in common by all imaginary time codes. We can conclude from ELK, FHI-AIMS, Kutepov et al, that 50 or so grid points should be sufficient, and VASP indicates that 20  grid points may also work. 
\item \textbf{Point group symmetry.} Taking advantage of point group symmetry, in solid state calculations, can easily give a factor of ten in memory and speed, by reducing  the number of kpoints.  I believe that some of the plane wave codes have not yet performed this optimization.
\item \textbf{Representation in position space.} In a plane wave code implementing the  space-time method, one must transform from momentum space to position space, multiply two operators, and then transform back to momentum space.   Since the basis size multiplies a number in the range $ \left[2,8\right]$, the total memory multiplies by $\times \left[4, 64 \right]$.  This memory issue must be controlled and optimized. I believe that the LBNL code did not actually address this issue and ended up with position space matrices that were a factor of $60$ too big. Godby's code did address the issue by minimizing the position space basis, which caused a loss of information.    I believe that  VASP handles the problem by generating the polarizability matrix one row at a time, and truncating each row after it has been transformed back into momentum space.   There are some other ways of handling this, including performing convolutions of $G$ (thus avoiding position space entirely, at a computational expense), and exploring ways of smearing the real space mesh.
\item \textbf{Reduction of the basis size.} In plane wave codes the very popular way of doing this is to use a reduced cutoff (compared to the one used in DFT) for the polarizability, susceptibility, screened potential, and self-energy.   I think that perhaps VASP uses this approach, while the LBNL code certainly does not.  I believe that Godby did not either.
\item \textbf{Band-oriented cutoffs.} VASP, like many of the non-plane-wave codes, has a cutoff limiting the number of bands used from the DFT result. However the cutoff is rather large, or in supercell calculations includes  many states, so VASP does a full diagonalization of the full DFT Hamiltonian, using Scalapack.  This sort of cutoff is not necessary if using iterative techniques like the Sternheimer approach, or if using full-matrix techniques.
\item \textbf{Parallelization.}  The parallelization strategies used by the various codes are not entirely clear.     In an imaginary time plane wave code, the first level of parallelization is to distribute kpoints, times, and spins.  Probably the second level is to parallelize over Gvectors.
\end{enumerate}
I believe that no plane wave code has completely implemented even this basic set of optimizations, so it is noteworthy that all of them report satisfactory errors and agreement with other publications, even while using supercells.

%

\begin{acknowledgments}
We gratefully acknowledge stimulating and in several cases pivotal discussions with Q. Wu, L. Du, T. Ozaki, M. J. Han, H. K. Yoon, S. U. Jang, S. H. Ryee, K. Refson, P. Hasnip, S. J. Clark, C. Weber, E. Plekhanov, J. Yates, M. Probert, D. Jochym, M. Lueders, R. Godby, J. Wetherell,  A. Davydov, A. Kim, G. Booth, M. Schilfgaarde, K. Haule,  S. Savrasov, F. Giustino, V. Olevano, X. Blase,   V. Antropov,   C. Carbogno, V. Blum, F. Manby,  S. Blugel, P. Werner, F. Aryasetiawan, S. Biermann,  P. Liu, and G. Kresse.  We acknowledge  support from EPSRC grant EP/M011038/1. 
  \end{acknowledgments}

\bibliography{Vincent}

\begin{thebibliography}{32}
\expandafter\ifx\csname natexlab\endcsname\relax\def\natexlab#1{#1}\fi
\expandafter\ifx\csname bibnamefont\endcsname\relax
  \def\bibnamefont#1{#1}\fi
\expandafter\ifx\csname bibfnamefont\endcsname\relax
  \def\bibfnamefont#1{#1}\fi
\expandafter\ifx\csname citenamefont\endcsname\relax
  \def\citenamefont#1{#1}\fi
\expandafter\ifx\csname url\endcsname\relax
  \def\url#1{\texttt{#1}}\fi
\expandafter\ifx\csname urlprefix\endcsname\relax\def\urlprefix{URL }\fi
\providecommand{\bibinfo}[2]{#2}
\providecommand{\eprint}[2][]{\url{#2}}

\bibitem[{\citenamefont{Kaltak et~al.}(2014{\natexlab{a}})\citenamefont{Kaltak,
  Klime¨, and Kresse}}]{kaltak2014low}
\bibinfo{author}{\bibfnamefont{M.}~\bibnamefont{Kaltak}},
  \bibinfo{author}{\bibfnamefont{J.}~\bibnamefont{Klime¨}}, \bibnamefont{and}
  \bibinfo{author}{\bibfnamefont{G.}~\bibnamefont{Kresse}},
  \bibinfo{journal}{Journal of chemical theory and computation}
  \textbf{\bibinfo{volume}{10}}, \bibinfo{pages}{2498}
  (\bibinfo{year}{2014}{\natexlab{a}}).

\bibitem[{\citenamefont{Kaltak et~al.}(2014{\natexlab{b}})\citenamefont{Kaltak,
  Klime\ifmmode~\check{s}\else \v{s}\fi{}, and Kresse}}]{PhysRevB.90.054115}
\bibinfo{author}{\bibfnamefont{M.}~\bibnamefont{Kaltak}},
  \bibinfo{author}{\bibfnamefont{J.~c.~v.}
  \bibnamefont{Klime\ifmmode~\check{s}\else \v{s}\fi{}}}, \bibnamefont{and}
  \bibinfo{author}{\bibfnamefont{G.}~\bibnamefont{Kresse}},
  \bibinfo{journal}{Phys. Rev. B} \textbf{\bibinfo{volume}{90}},
  \bibinfo{pages}{054115} (\bibinfo{year}{2014}{\natexlab{b}}),
  \urlprefix\url{http://link.aps.org/doi/10.1103/PhysRevB.90.054115}.

\bibitem[{\citenamefont{Liu et~al.}(2016)\citenamefont{Liu, Kaltak,
  Klime\ifmmode~\check{s}\else \v{s}\fi{}, and Kresse}}]{PhysRevB.94.165109}
\bibinfo{author}{\bibfnamefont{P.}~\bibnamefont{Liu}},
  \bibinfo{author}{\bibfnamefont{M.}~\bibnamefont{Kaltak}},
  \bibinfo{author}{\bibfnamefont{J.~c.~v.}
  \bibnamefont{Klime\ifmmode~\check{s}\else \v{s}\fi{}}}, \bibnamefont{and}
  \bibinfo{author}{\bibfnamefont{G.}~\bibnamefont{Kresse}},
  \bibinfo{journal}{Phys. Rev. B} \textbf{\bibinfo{volume}{94}},
  \bibinfo{pages}{165109} (\bibinfo{year}{2016}),
  \urlprefix\url{http://link.aps.org/doi/10.1103/PhysRevB.94.165109}.

\bibitem[{\citenamefont{Wang}(2015)}]{PhysRevB.91.125135}
\bibinfo{author}{\bibfnamefont{L.-W.} \bibnamefont{Wang}},
  \bibinfo{journal}{Phys. Rev. B} \textbf{\bibinfo{volume}{91}},
  \bibinfo{pages}{125135} (\bibinfo{year}{2015}),
  \urlprefix\url{http://link.aps.org/doi/10.1103/PhysRevB.91.125135}.

\bibitem[{\citenamefont{Rojas et~al.}(1995)\citenamefont{Rojas, Godby, and
  Needs}}]{PhysRevLett.74.1827}
\bibinfo{author}{\bibfnamefont{H.~N.} \bibnamefont{Rojas}},
  \bibinfo{author}{\bibfnamefont{R.~W.} \bibnamefont{Godby}}, \bibnamefont{and}
  \bibinfo{author}{\bibfnamefont{R.~J.} \bibnamefont{Needs}},
  \bibinfo{journal}{Phys. Rev. Lett.} \textbf{\bibinfo{volume}{74}},
  \bibinfo{pages}{1827} (\bibinfo{year}{1995}),
  \urlprefix\url{http://link.aps.org/doi/10.1103/PhysRevLett.74.1827}.

\bibitem[{\citenamefont{Rieger et~al.}(1999)\citenamefont{Rieger, Steinbeck,
  White, Rojas, and Godby}}]{rieger1999gw}
\bibinfo{author}{\bibfnamefont{M.~M.} \bibnamefont{Rieger}},
  \bibinfo{author}{\bibfnamefont{L.}~\bibnamefont{Steinbeck}},
  \bibinfo{author}{\bibfnamefont{I.}~\bibnamefont{White}},
  \bibinfo{author}{\bibfnamefont{H.}~\bibnamefont{Rojas}}, \bibnamefont{and}
  \bibinfo{author}{\bibfnamefont{R.}~\bibnamefont{Godby}},
  \bibinfo{journal}{Computer physics communications}
  \textbf{\bibinfo{volume}{117}}, \bibinfo{pages}{211} (\bibinfo{year}{1999}).

\bibitem[{\citenamefont{Steinbeck et~al.}(2000)\citenamefont{Steinbeck, Rubio,
  Reining, Torrent, White, and Godby}}]{Steinbeck2000}
\bibinfo{author}{\bibfnamefont{L.}~\bibnamefont{Steinbeck}},
  \bibinfo{author}{\bibfnamefont{A.}~\bibnamefont{Rubio}},
  \bibinfo{author}{\bibfnamefont{L.}~\bibnamefont{Reining}},
  \bibinfo{author}{\bibfnamefont{M.}~\bibnamefont{Torrent}},
  \bibinfo{author}{\bibfnamefont{I.}~\bibnamefont{White}}, \bibnamefont{and}
  \bibinfo{author}{\bibfnamefont{R.}~\bibnamefont{Godby}},
  \bibinfo{journal}{Computer Physics Communications}
  \textbf{\bibinfo{volume}{125}}, \bibinfo{pages}{105} (\bibinfo{year}{2000}).

\bibitem[{\citenamefont{Chu et~al.}(2016)\citenamefont{Chu, Trinastic, Wang,
  Eguiluz, Kozhevnikov, Schulthess, and Cheng}}]{PhysRevB.93.125210}
\bibinfo{author}{\bibfnamefont{I.-H.} \bibnamefont{Chu}},
  \bibinfo{author}{\bibfnamefont{J.~P.} \bibnamefont{Trinastic}},
  \bibinfo{author}{\bibfnamefont{Y.-P.} \bibnamefont{Wang}},
  \bibinfo{author}{\bibfnamefont{A.~G.} \bibnamefont{Eguiluz}},
  \bibinfo{author}{\bibfnamefont{A.}~\bibnamefont{Kozhevnikov}},
  \bibinfo{author}{\bibfnamefont{T.~C.} \bibnamefont{Schulthess}},
  \bibnamefont{and} \bibinfo{author}{\bibfnamefont{H.-P.} \bibnamefont{Cheng}},
  \bibinfo{journal}{Phys. Rev. B} \textbf{\bibinfo{volume}{93}},
  \bibinfo{pages}{125210} (\bibinfo{year}{2016}),
  \urlprefix\url{http://link.aps.org/doi/10.1103/PhysRevB.93.125210}.

\bibitem[{\citenamefont{Ren et~al.}(2009)\citenamefont{Ren, Rinke, and
  Scheffler}}]{PhysRevB.80.045402}
\bibinfo{author}{\bibfnamefont{X.}~\bibnamefont{Ren}},
  \bibinfo{author}{\bibfnamefont{P.}~\bibnamefont{Rinke}}, \bibnamefont{and}
  \bibinfo{author}{\bibfnamefont{M.}~\bibnamefont{Scheffler}},
  \bibinfo{journal}{Phys. Rev. B} \textbf{\bibinfo{volume}{80}},
  \bibinfo{pages}{045402} (\bibinfo{year}{2009}),
  \urlprefix\url{http://link.aps.org/doi/10.1103/PhysRevB.80.045402}.

\bibitem[{\citenamefont{van Setten et~al.}(2015)\citenamefont{van Setten,
  Caruso, Sharifzadeh, Ren, Scheffler, Liu, Lischner, Lin, Deslippe, Louie
  et~al.}}]{van2015gw}
\bibinfo{author}{\bibfnamefont{M.~J.} \bibnamefont{van Setten}},
  \bibinfo{author}{\bibfnamefont{F.}~\bibnamefont{Caruso}},
  \bibinfo{author}{\bibfnamefont{S.}~\bibnamefont{Sharifzadeh}},
  \bibinfo{author}{\bibfnamefont{X.}~\bibnamefont{Ren}},
  \bibinfo{author}{\bibfnamefont{M.}~\bibnamefont{Scheffler}},
  \bibinfo{author}{\bibfnamefont{F.}~\bibnamefont{Liu}},
  \bibinfo{author}{\bibfnamefont{J.}~\bibnamefont{Lischner}},
  \bibinfo{author}{\bibfnamefont{L.}~\bibnamefont{Lin}},
  \bibinfo{author}{\bibfnamefont{J.~R.} \bibnamefont{Deslippe}},
  \bibinfo{author}{\bibfnamefont{S.~G.} \bibnamefont{Louie}},
  \bibnamefont{et~al.}, \bibinfo{journal}{Journal of chemical theory and
  computation} \textbf{\bibinfo{volume}{11}}, \bibinfo{pages}{5665}
  (\bibinfo{year}{2015}).

\bibitem[{\citenamefont{Ren et~al.}(2015)\citenamefont{Ren, Marom, Caruso,
  Scheffler, and Rinke}}]{PhysRevB.92.081104}
\bibinfo{author}{\bibfnamefont{X.}~\bibnamefont{Ren}},
  \bibinfo{author}{\bibfnamefont{N.}~\bibnamefont{Marom}},
  \bibinfo{author}{\bibfnamefont{F.}~\bibnamefont{Caruso}},
  \bibinfo{author}{\bibfnamefont{M.}~\bibnamefont{Scheffler}},
  \bibnamefont{and} \bibinfo{author}{\bibfnamefont{P.}~\bibnamefont{Rinke}},
  \bibinfo{journal}{Phys. Rev. B} \textbf{\bibinfo{volume}{92}},
  \bibinfo{pages}{081104} (\bibinfo{year}{2015}),
  \urlprefix\url{http://link.aps.org/doi/10.1103/PhysRevB.92.081104}.

\bibitem[{\citenamefont{Caruso et~al.}(2013{\natexlab{a}})\citenamefont{Caruso,
  Rohr, Hellgren, Ren, Rinke, Rubio, and Scheffler}}]{PhysRevLett.110.146403}
\bibinfo{author}{\bibfnamefont{F.}~\bibnamefont{Caruso}},
  \bibinfo{author}{\bibfnamefont{D.~R.} \bibnamefont{Rohr}},
  \bibinfo{author}{\bibfnamefont{M.}~\bibnamefont{Hellgren}},
  \bibinfo{author}{\bibfnamefont{X.}~\bibnamefont{Ren}},
  \bibinfo{author}{\bibfnamefont{P.}~\bibnamefont{Rinke}},
  \bibinfo{author}{\bibfnamefont{A.}~\bibnamefont{Rubio}}, \bibnamefont{and}
  \bibinfo{author}{\bibfnamefont{M.}~\bibnamefont{Scheffler}},
  \bibinfo{journal}{Phys. Rev. Lett.} \textbf{\bibinfo{volume}{110}},
  \bibinfo{pages}{146403} (\bibinfo{year}{2013}{\natexlab{a}}),
  \urlprefix\url{http://link.aps.org/doi/10.1103/PhysRevLett.110.146403}.

\bibitem[{\citenamefont{Ren et~al.}(2013)\citenamefont{Ren, Rinke, Scuseria,
  and Scheffler}}]{PhysRevB.88.035120}
\bibinfo{author}{\bibfnamefont{X.}~\bibnamefont{Ren}},
  \bibinfo{author}{\bibfnamefont{P.}~\bibnamefont{Rinke}},
  \bibinfo{author}{\bibfnamefont{G.~E.} \bibnamefont{Scuseria}},
  \bibnamefont{and}
  \bibinfo{author}{\bibfnamefont{M.}~\bibnamefont{Scheffler}},
  \bibinfo{journal}{Phys. Rev. B} \textbf{\bibinfo{volume}{88}},
  \bibinfo{pages}{035120} (\bibinfo{year}{2013}),
  \urlprefix\url{http://link.aps.org/doi/10.1103/PhysRevB.88.035120}.

\bibitem[{\citenamefont{Caruso et~al.}(2013{\natexlab{b}})\citenamefont{Caruso,
  Rinke, Ren, Rubio, and Scheffler}}]{PhysRevB.88.075105}
\bibinfo{author}{\bibfnamefont{F.}~\bibnamefont{Caruso}},
  \bibinfo{author}{\bibfnamefont{P.}~\bibnamefont{Rinke}},
  \bibinfo{author}{\bibfnamefont{X.}~\bibnamefont{Ren}},
  \bibinfo{author}{\bibfnamefont{A.}~\bibnamefont{Rubio}}, \bibnamefont{and}
  \bibinfo{author}{\bibfnamefont{M.}~\bibnamefont{Scheffler}},
  \bibinfo{journal}{Phys. Rev. B} \textbf{\bibinfo{volume}{88}},
  \bibinfo{pages}{075105} (\bibinfo{year}{2013}{\natexlab{b}}),
  \urlprefix\url{http://link.aps.org/doi/10.1103/PhysRevB.88.075105}.

\bibitem[{\citenamefont{Caruso et~al.}(2012)\citenamefont{Caruso, Rinke, Ren,
  Scheffler, and Rubio}}]{PhysRevB.86.081102}
\bibinfo{author}{\bibfnamefont{F.}~\bibnamefont{Caruso}},
  \bibinfo{author}{\bibfnamefont{P.}~\bibnamefont{Rinke}},
  \bibinfo{author}{\bibfnamefont{X.}~\bibnamefont{Ren}},
  \bibinfo{author}{\bibfnamefont{M.}~\bibnamefont{Scheffler}},
  \bibnamefont{and} \bibinfo{author}{\bibfnamefont{A.}~\bibnamefont{Rubio}},
  \bibinfo{journal}{Phys. Rev. B} \textbf{\bibinfo{volume}{86}},
  \bibinfo{pages}{081102} (\bibinfo{year}{2012}),
  \urlprefix\url{http://link.aps.org/doi/10.1103/PhysRevB.86.081102}.

\bibitem[{\citenamefont{Paier et~al.}(2012)\citenamefont{Paier, Ren, Rinke,
  Scuseria, Gr{\"u}neis, Kresse, and Scheffler}}]{paier2012assessment}
\bibinfo{author}{\bibfnamefont{J.}~\bibnamefont{Paier}},
  \bibinfo{author}{\bibfnamefont{X.}~\bibnamefont{Ren}},
  \bibinfo{author}{\bibfnamefont{P.}~\bibnamefont{Rinke}},
  \bibinfo{author}{\bibfnamefont{G.~E.} \bibnamefont{Scuseria}},
  \bibinfo{author}{\bibfnamefont{A.}~\bibnamefont{Gr{\"u}neis}},
  \bibinfo{author}{\bibfnamefont{G.}~\bibnamefont{Kresse}}, \bibnamefont{and}
  \bibinfo{author}{\bibfnamefont{M.}~\bibnamefont{Scheffler}},
  \bibinfo{journal}{New Journal of Physics} \textbf{\bibinfo{volume}{14}},
  \bibinfo{pages}{043002} (\bibinfo{year}{2012}).

\bibitem[{\citenamefont{Kutepov et~al.}(2012)\citenamefont{Kutepov, Haule,
  Savrasov, and Kotliar}}]{PhysRevB.85.155129}
\bibinfo{author}{\bibfnamefont{A.}~\bibnamefont{Kutepov}},
  \bibinfo{author}{\bibfnamefont{K.}~\bibnamefont{Haule}},
  \bibinfo{author}{\bibfnamefont{S.~Y.} \bibnamefont{Savrasov}},
  \bibnamefont{and} \bibinfo{author}{\bibfnamefont{G.}~\bibnamefont{Kotliar}},
  \bibinfo{journal}{Phys. Rev. B} \textbf{\bibinfo{volume}{85}},
  \bibinfo{pages}{155129} (\bibinfo{year}{2012}),
  \urlprefix\url{http://link.aps.org/doi/10.1103/PhysRevB.85.155129}.

\bibitem[{\citenamefont{Kutepov et~al.}(2009)\citenamefont{Kutepov, Savrasov,
  and Kotliar}}]{PhysRevB.80.041103}
\bibinfo{author}{\bibfnamefont{A.}~\bibnamefont{Kutepov}},
  \bibinfo{author}{\bibfnamefont{S.~Y.} \bibnamefont{Savrasov}},
  \bibnamefont{and} \bibinfo{author}{\bibfnamefont{G.}~\bibnamefont{Kotliar}},
  \bibinfo{journal}{Phys. Rev. B} \textbf{\bibinfo{volume}{80}},
  \bibinfo{pages}{041103} (\bibinfo{year}{2009}),
  \urlprefix\url{http://link.aps.org/doi/10.1103/PhysRevB.80.041103}.

\bibitem[{\citenamefont{Zein et~al.}(2006)\citenamefont{Zein, Savrasov, and
  Kotliar}}]{PhysRevLett.96.226403}
\bibinfo{author}{\bibfnamefont{N.~E.} \bibnamefont{Zein}},
  \bibinfo{author}{\bibfnamefont{S.~Y.} \bibnamefont{Savrasov}},
  \bibnamefont{and} \bibinfo{author}{\bibfnamefont{G.}~\bibnamefont{Kotliar}},
  \bibinfo{journal}{Phys. Rev. Lett.} \textbf{\bibinfo{volume}{96}},
  \bibinfo{pages}{226403} (\bibinfo{year}{2006}),
  \urlprefix\url{http://link.aps.org/doi/10.1103/PhysRevLett.96.226403}.

\bibitem[{\citenamefont{Choi et~al.}(2015)\citenamefont{Choi, Kutepov, Haule,
  van Schilfgaarde, and Kotliar}}]{choi2015first}
\bibinfo{author}{\bibfnamefont{S.}~\bibnamefont{Choi}},
  \bibinfo{author}{\bibfnamefont{A.}~\bibnamefont{Kutepov}},
  \bibinfo{author}{\bibfnamefont{K.}~\bibnamefont{Haule}},
  \bibinfo{author}{\bibfnamefont{M.}~\bibnamefont{van Schilfgaarde}},
  \bibnamefont{and} \bibinfo{author}{\bibfnamefont{G.}~\bibnamefont{Kotliar}},
  \bibinfo{journal}{arXiv preprint arXiv:1504.07569}  (\bibinfo{year}{2015}).

\bibitem[{\citenamefont{Sponza et~al.}(2016)\citenamefont{Sponza, Pisanti,
  Vishina, Pashov, Vidal, Weber, Kotliar, and van
  Schilfgaarde}}]{sponza2016non}
\bibinfo{author}{\bibfnamefont{L.}~\bibnamefont{Sponza}},
  \bibinfo{author}{\bibfnamefont{P.}~\bibnamefont{Pisanti}},
  \bibinfo{author}{\bibfnamefont{A.}~\bibnamefont{Vishina}},
  \bibinfo{author}{\bibfnamefont{D.}~\bibnamefont{Pashov}},
  \bibinfo{author}{\bibfnamefont{J.}~\bibnamefont{Vidal}},
  \bibinfo{author}{\bibfnamefont{C.}~\bibnamefont{Weber}},
  \bibinfo{author}{\bibfnamefont{G.}~\bibnamefont{Kotliar}}, \bibnamefont{and}
  \bibinfo{author}{\bibfnamefont{M.}~\bibnamefont{van Schilfgaarde}},
  \bibinfo{journal}{arXiv preprint arXiv:1603.05521}  (\bibinfo{year}{2016}).

\bibitem[{\citenamefont{Koval et~al.}(2014)\citenamefont{Koval, Foerster, and
  S\'anchez-Portal}}]{PhysRevB.89.155417}
\bibinfo{author}{\bibfnamefont{P.}~\bibnamefont{Koval}},
  \bibinfo{author}{\bibfnamefont{D.}~\bibnamefont{Foerster}}, \bibnamefont{and}
  \bibinfo{author}{\bibfnamefont{D.}~\bibnamefont{S\'anchez-Portal}},
  \bibinfo{journal}{Phys. Rev. B} \textbf{\bibinfo{volume}{89}},
  \bibinfo{pages}{155417} (\bibinfo{year}{2014}),
  \urlprefix\url{http://link.aps.org/doi/10.1103/PhysRevB.89.155417}.

\bibitem[{\citenamefont{Foerster et~al.}(2011)\citenamefont{Foerster, Koval,
  and S{\'a}nchez-Portal}}]{foerster2011n3}
\bibinfo{author}{\bibfnamefont{D.}~\bibnamefont{Foerster}},
  \bibinfo{author}{\bibfnamefont{P.}~\bibnamefont{Koval}}, \bibnamefont{and}
  \bibinfo{author}{\bibfnamefont{D.}~\bibnamefont{S{\'a}nchez-Portal}},
  \bibinfo{journal}{The Journal of chemical physics}
  \textbf{\bibinfo{volume}{135}}, \bibinfo{pages}{074105}
  (\bibinfo{year}{2011}).

\bibitem[{\citenamefont{Rostgaard et~al.}(2010)\citenamefont{Rostgaard,
  Jacobsen, and Thygesen}}]{PhysRevB.81.085103}
\bibinfo{author}{\bibfnamefont{C.}~\bibnamefont{Rostgaard}},
  \bibinfo{author}{\bibfnamefont{K.~W.} \bibnamefont{Jacobsen}},
  \bibnamefont{and} \bibinfo{author}{\bibfnamefont{K.~S.}
  \bibnamefont{Thygesen}}, \bibinfo{journal}{Phys. Rev. B}
  \textbf{\bibinfo{volume}{81}}, \bibinfo{pages}{085103}
  (\bibinfo{year}{2010}),
  \urlprefix\url{http://link.aps.org/doi/10.1103/PhysRevB.81.085103}.

\bibitem[{\citenamefont{Strange et~al.}(2011)\citenamefont{Strange, Rostgaard,
  H\"akkinen, and Thygesen}}]{PhysRevB.83.115108}
\bibinfo{author}{\bibfnamefont{M.}~\bibnamefont{Strange}},
  \bibinfo{author}{\bibfnamefont{C.}~\bibnamefont{Rostgaard}},
  \bibinfo{author}{\bibfnamefont{H.}~\bibnamefont{H\"akkinen}},
  \bibnamefont{and} \bibinfo{author}{\bibfnamefont{K.~S.}
  \bibnamefont{Thygesen}}, \bibinfo{journal}{Phys. Rev. B}
  \textbf{\bibinfo{volume}{83}}, \bibinfo{pages}{115108}
  (\bibinfo{year}{2011}),
  \urlprefix\url{http://link.aps.org/doi/10.1103/PhysRevB.83.115108}.

\bibitem[{\citenamefont{Stan et~al.}(2006)\citenamefont{Stan, Dahlen, and
  Van~Leeuwen}}]{stan2006fully}
\bibinfo{author}{\bibfnamefont{A.}~\bibnamefont{Stan}},
  \bibinfo{author}{\bibfnamefont{N.~E.} \bibnamefont{Dahlen}},
  \bibnamefont{and}
  \bibinfo{author}{\bibfnamefont{R.}~\bibnamefont{Van~Leeuwen}},
  \bibinfo{journal}{EPL (Europhysics Letters)} \textbf{\bibinfo{volume}{76}},
  \bibinfo{pages}{298} (\bibinfo{year}{2006}).

\bibitem[{\citenamefont{Dahlen and van Leeuwen}(2005)}]{dahlen2005self}
\bibinfo{author}{\bibfnamefont{N.~E.} \bibnamefont{Dahlen}} \bibnamefont{and}
  \bibinfo{author}{\bibfnamefont{R.}~\bibnamefont{van Leeuwen}},
  \bibinfo{journal}{The Journal of chemical physics}
  \textbf{\bibinfo{volume}{122}}, \bibinfo{pages}{164102}
  (\bibinfo{year}{2005}).

\bibitem[{\citenamefont{Stan et~al.}(2009)\citenamefont{Stan, Dahlen, and
  Van~Leeuwen}}]{stan2009levels}
\bibinfo{author}{\bibfnamefont{A.}~\bibnamefont{Stan}},
  \bibinfo{author}{\bibfnamefont{N.~E.} \bibnamefont{Dahlen}},
  \bibnamefont{and}
  \bibinfo{author}{\bibfnamefont{R.}~\bibnamefont{Van~Leeuwen}},
  \bibinfo{journal}{J. Chem. Phys.} \textbf{\bibinfo{volume}{130}},
  \bibinfo{pages}{114105} (\bibinfo{year}{2009}).

\bibitem[{\citenamefont{Zein and Antropov}(2002)}]{PhysRevLett.89.126402}
\bibinfo{author}{\bibfnamefont{N.~E.} \bibnamefont{Zein}} \bibnamefont{and}
  \bibinfo{author}{\bibfnamefont{V.~P.} \bibnamefont{Antropov}},
  \bibinfo{journal}{Phys. Rev. Lett.} \textbf{\bibinfo{volume}{89}},
  \bibinfo{pages}{126402} (\bibinfo{year}{2002}),
  \urlprefix\url{http://link.aps.org/doi/10.1103/PhysRevLett.89.126402}.

\bibitem[{\citenamefont{Ku and Eguiluz}(2002)}]{PhysRevLett.89.126401}
\bibinfo{author}{\bibfnamefont{W.}~\bibnamefont{Ku}} \bibnamefont{and}
  \bibinfo{author}{\bibfnamefont{A.~G.} \bibnamefont{Eguiluz}},
  \bibinfo{journal}{Phys. Rev. Lett.} \textbf{\bibinfo{volume}{89}},
  \bibinfo{pages}{126401} (\bibinfo{year}{2002}),
  \urlprefix\url{http://link.aps.org/doi/10.1103/PhysRevLett.89.126401}.

\bibitem[{\citenamefont{Sch\"one and Eguiluz}(1998)}]{PhysRevLett.81.1662}
\bibinfo{author}{\bibfnamefont{W.-D.} \bibnamefont{Sch\"one}} \bibnamefont{and}
  \bibinfo{author}{\bibfnamefont{A.~G.} \bibnamefont{Eguiluz}},
  \bibinfo{journal}{Phys. Rev. Lett.} \textbf{\bibinfo{volume}{81}},
  \bibinfo{pages}{1662} (\bibinfo{year}{1998}),
  \urlprefix\url{http://link.aps.org/doi/10.1103/PhysRevLett.81.1662}.

\bibitem[{\citenamefont{Eguiluz and Schone}(1998)}]{eguiluz1998electronic}
\bibinfo{author}{\bibfnamefont{A.~G.} \bibnamefont{Eguiluz}} \bibnamefont{and}
  \bibinfo{author}{\bibfnamefont{W.-D.} \bibnamefont{Schone}},
  \bibinfo{journal}{Molecular Physics} \textbf{\bibinfo{volume}{94}},
  \bibinfo{pages}{87} (\bibinfo{year}{1998}).

\end{thebibliography}

\end{document}